\documentclass[aps,prd,showpacs,nobibnotes,twocolumn,preprintnumbers,
nobalancelastpage,superscriptaddress]{revtex4}
\usepackage{latexsym}
\usepackage{dcolumn}
\usepackage{bm}
\usepackage{amsmath}
\usepackage{natbib}
\input{epsf}
\newcommand{\be}{\begin{equation}}
\newcommand{\ee}{\end{equation}}
\newcommand{\ba}{\begin{eqnarray}}
\newcommand{\ea}{\end{eqnarray}}
\newcommand{\bi}{\begin{itemize}}
\newcommand{\ei}{\end{itemize}}
\newcommand{\bfi}{\begin{figure}[!t]
\epsfxsize=9cm
\epsffile}
\newcommand{\bfig}{\begin{figure*}[t]
\epsfxsize=15cm
\epsffile}
\newcommand{\efi}{\end{figure}}
\newcommand{\efig}{\end{figure*}}

\newcommand{\la}{\lesssim}
\newcommand{\ga}{\gtrsim}

\newcommand{\hmpc}{h/{\rm Mpc}}
\begin{document}
\title{The large scale structure in the 3D luminosity-distance space and its cosmological applications}
\author{Pengjie Zhang}
\affiliation{Department of Astronomy, School of Physics and Astronomy, Shanghai Jiao Tong
  University, Shanghai, 200240, China}
\email[Email me at: ]{zhangpj@sjtu.edu.cn}
\affiliation{Tsung-Dao Lee institute, Shanghai, 200240, China}
\affiliation{IFSA Collaborative Innovation Center, Shanghai Jiao Tong
University, Shanghai 200240, China}
\affiliation{Shanghai Key Laboratory for Particle Physics and Cosmology, Shanghai 200240, China}
\begin{abstract}
Future gravitational wave (GW) observations are capable of detecting
millions of compact star binary mergers in
extragalactic galaxies, with $1\%$ luminosity-distance ($D_L$) measurement accuracy and
better than arcminute positioning accuracy. This
will open a new window of the large scale structure (LSS) of the
universe, in the 3D {\bf luminosity-distance space (LDS)}, instead of the 3D
redshift space of galaxy spectroscopic surveys. The baryon acoustic oscillation and the AP
test encoded in the  LDS LSS  constrain the $D_L$-$D^{\rm co}_A$ (comoving angular
diameter distance) relation and therefore the expansion history of the
universe. Peculiar velocity induces the LDS distortion, analogous to the redshift space distortion, and allows for
a new structure growth measure $f_L\sigma_8$. When the distance
duality is enforced ($1+z=D_L/D^{\rm co}_A$), the LDS LSS by itself determines the redshift to
$\sim 1\%$ level accuracy, 
and alleviates the need of spectroscopic follow-up of GW
events.But a more valuable application is to test the distance duality to
$1\%$ level accuracy, in combination with conventional BAO and 
supernovae measurements. This will put stringent constraints on modified
gravity models in which the gravitational wave $D^{\rm GW}_L$
deviates from the electromagnetic wave $D^{EM}_L$. All these applications
require no spectroscopic follow-ups. 
\end{abstract}
\maketitle


{\bf Introduction}.---
Discoveries of gravitational wave (GW) produced by black hole
(BH)/neutron star (NS)-BH/NS mergers
\citep{2016PhRvL.116f1102A,2017PhRvL.118v1101A,2017PhRvL.119n1101A,2017PhRvL.119p1101A}
have opened the era of  
gravitational wave astronomy. These GW events can serve as standard
sirens to measure cosmological distance from first principles
\citep{1986Natur.323..310S,2017Natur.551...85A} and therefore  avoid
various systematics associated with traditional 
methods. It will then have profound impact on
cosmology. However, to fulfill this potential, usually it requires
spectroscopic follow-ups to determine redshifts of their host
galaxies or electromagnetic counterparts. This will be challenging, for
the third generation GW experiments such as the Big Bang Observer
(BBO, \cite{2006PhRvD..73d2001C, 2009PhRvD..80j4009C}), and the
Einstein Telescope\footnote{https://tds.virgo-gw.eu/?call\_file=ET\-0106C\-10.pdf}, which will detect millions of
these GW events.  Various alternatives have been proposed to
circumvent this stringent need of spectroscopic follow-ups \cite{2016PhRvL.116l1302N,2016PhRvD..93h3511O,2018PhRvD..98b3502N,2018arXiv180806615M}. 

We point out a new possibility to circumvent this challenge. These GW
events are hosted by galaxies and are 
therefore tracers of the large scale structure (LSS). With arcminute
positioning accuracy and $1\%$ level accuracy in the luminosity distance
$D_L$ determination achievable by BBO,  we are able to map the {\bf 3D large scale
structure in the luminosity-distance space (LDS)}.  It is analogous to
the redshift space LSS mapped by the conventional spectroscopic
redshift surveys of galaxies ($D_L\leftrightarrow z$). Therefore it also contains valuable 
information of baryon acoustic oscillation (BAO), both across the sky
and along the line of sight. As BAO in the redshift space measures the
comoving angular diameter distance
$D_A^{\rm co}$ and $H(z)=dz/d\chi$ at given redshift bins, BAO in LDS
measures $D_A^{\rm co}$ and $H_L\equiv dD_L/d\chi$ at given $D_L$ bins. Here $\chi$ is
the comoving radial distance.  Both the $D_L$-$D_A^{\rm co}$ relation
and the $D_L$-$H_L$ relation constrain cosmology (Fig. \ref{fig:DL}), without the need of
redshift. Furthermore,  both $D_A^{\rm co}$ and $H_L$ can be converted
into cosmological redshift through the distance duality relation
$1+z=D_L/D_A^{\rm co}$. Similar to the redshift space distortion (RSD), peculiar velocity
also induces the luminosity-distance space distortion (LDSD). This will
enable a new measure of structure growth rate $f_L\sigma_8$, which
differs from $f\sigma_8$ measured in RSD  by a redshift dependent factor. 

\bfi{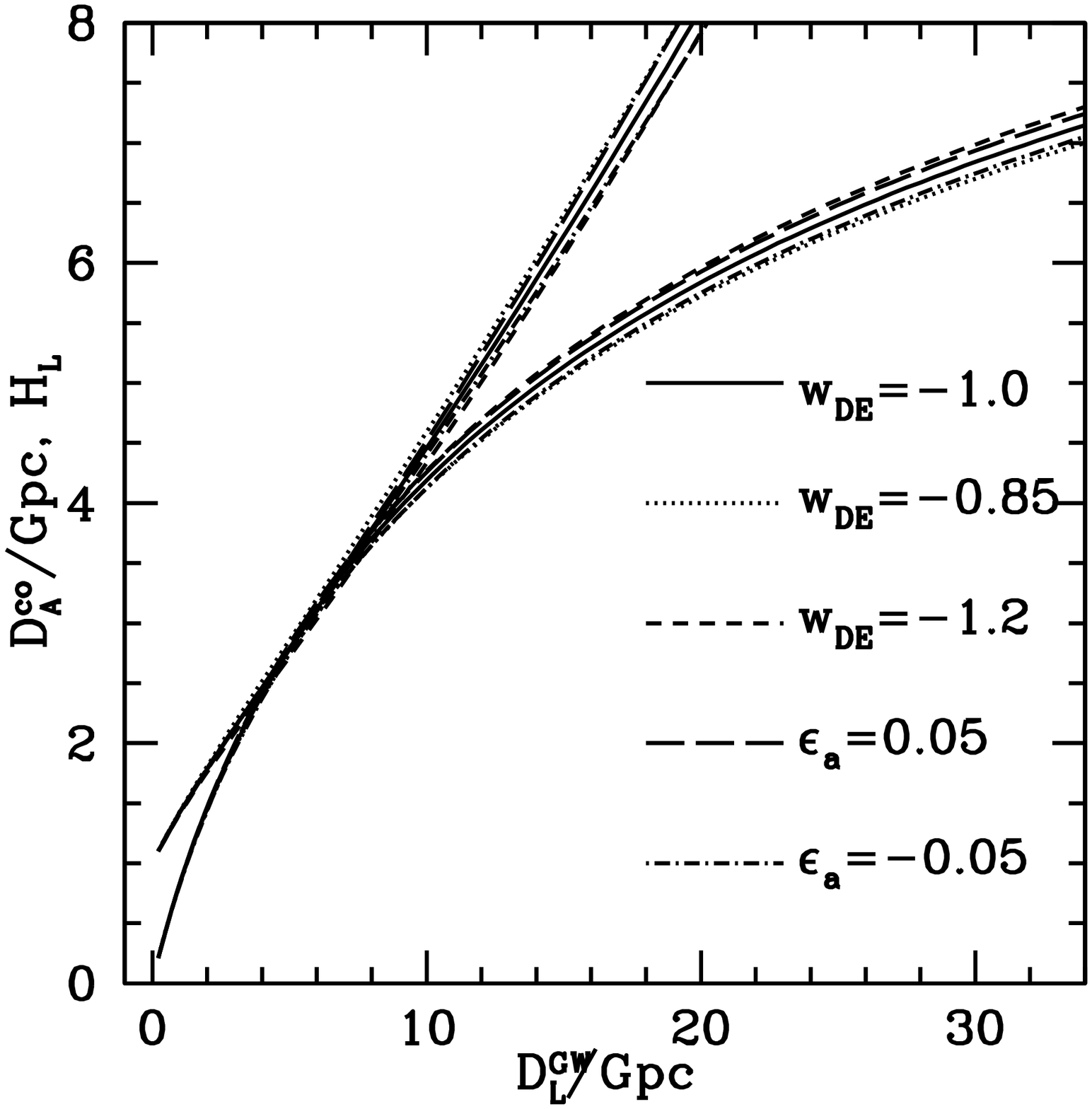}
\caption{The $D^{\rm GW}_L$-$D_A^{\rm co}$ and $D^{\rm GW}_L$-$H_L$ relations. Like
  the $z$-$D_A^{\rm co}$ relation and $z$-$H$ relations constrained
  by galaxy spectroscopic redshift surveys, the new set of relations
  is also sensitive to dark energy, demonstrated by the cases of
  various dark
  energy equation of state. A more
  unique application of these relations is to constrain
  modified gravity models in which $D^{\rm GW}_L\neq D^{\rm
    EM}_L$. We show two such cases, parameterized by $\epsilon_a=\pm
  0.05$. Using GW alone, $w_{\rm DE}$ and $\epsilon_a$ constraints
  are largely degenerate. Combination witt electromagnetic wave
  observations can break  this degeneracy straightforwardly.  \label{fig:DL}} 
\efi

{\bf The luminosity-distance space LSS}.---
Each GW event provides a 3D position ($D_L^{\rm
  obs}$, $\hat{n}$).  With millions of them, arcminute 
positioning accuracy, and $\mathcal{O}(1\%)$ accuracy in $D_L$,  we are able to
measure the number density fluctuation $\delta_{\rm GW}$ over
effectively the entire cosmic volume. This LSS is statistically anisotropic, since $D_L^{\rm
  obs}$ differs from its cosmological value $D_L$,
\ba
\label{eqn:DL}
D_L^{\rm obs}=D_L(1+2{\bf v}\cdot \hat{n}-\kappa +\cdots)\ .
\ea
Here $\kappa$ is the lensing convergence, describing the effect of
gravitational lensing magnification. This effect is a highly valuable
source of cosmological information 
(e.g. \cite{2009PhRvD..80j4009C}).  $\bf v$ is the physical
peculiar velocity \citep{2006PhRvD..73l3526H} and $\hat{n}$ is the
line of sight unit vector. If an object is moving
away from us (${\bf v}\cdot \hat{n}>0$), photons/GWs take longer time to reach us and suffer
more cosmic dimming. At $z\ga 1$, $\kappa\sim
\mathcal{O}(10^{-2})$ and ${\bf v}\cdot\hat{n}\sim
\mathcal{O}(10^{-3})$. Naively one would think the lensing effect
overwhelms the peculiar velocity effect. This is indeed the case if
we can subtract $D_L$ with cosmological redshift from spectroscopic follow-up. However, what affects the LDS LSS is the gradient of $\kappa$ and ${\bf v}\cdot \hat{n}$ along
the line of sight. Under the distance observer approximation and up to
leading order, $\delta^{\rm LDS}_{\rm GW}\simeq\delta_{\rm
  GW}+\alpha \nabla\kappa \cdot \hat{n}+\beta \nabla({\bf v}\cdot\hat{n}) \cdot \hat{n}$. Since $\kappa$ is lack of variation along the line of
sight, its contribution is sub-dominant comparing to the velocity
gradient contribution\footnote{The area amplification of lensing
  adds an extra term $-2\kappa$ to $\delta_{\rm GW}$. However, it is
  orders of magnitude smaller than the intrinsic clustering
  (e.g. \cite{YangXJ11}). }. The LDS  power spectrum then resembles
the Kaiser \cite{Kaiser87} plus Finger of God formula in RSD,
\ba
\label{eqn:LDSD}
P^{\rm LDS}(k_\perp,k_\parallel)=P_g(k)\left(1+\frac{f_L}{b_g}u^2\right)^2F(k_\parallel)\ .
\ea
Here $u\equiv k_\parallel/k$ and $k\equiv
\sqrt{k^2_\perp+k^2_\parallel}$. $k_\perp$ ($k_\parallel$) is the
wavevector perpendicular (parallel) to the line of sight. $b_g$ is the density bias of GW host
galaxies. $F(k_\parallel)$ describes the FOG
effect. There are two majo differences to RSD. First, 
\ba
\label{eqn:fL}
 f_{L}\equiv \left(\frac{2D_L/(1+z)}{d(D_L)/dz}\right)\times f\ .
\ea
It differs from $f\equiv d\ln D/d\ln a$ in RSD by a redshift dependent
factor. 
This arises from the different effects of peculiar velocity on the
luminosity distance  ($D_L\rightarrow D_L(1+2{\bf v}\cdot\hat{n}))$,
and on redshift ($z\rightarrow z+{\bf v}\cdot\hat{n}(1+z)$). The
prefactor in Eq. \ref{eqn:fL} is zero at $z=0$ and  increases with
$z$. It becomes larger than unity at $z\ga 1.7$, where the peculiar
velocity induced distortion is larger in LDS
than in redshift space.  The second difference is that the $H$ factor
shown up in FOG should be replaced by $H_L$. 

\bfi{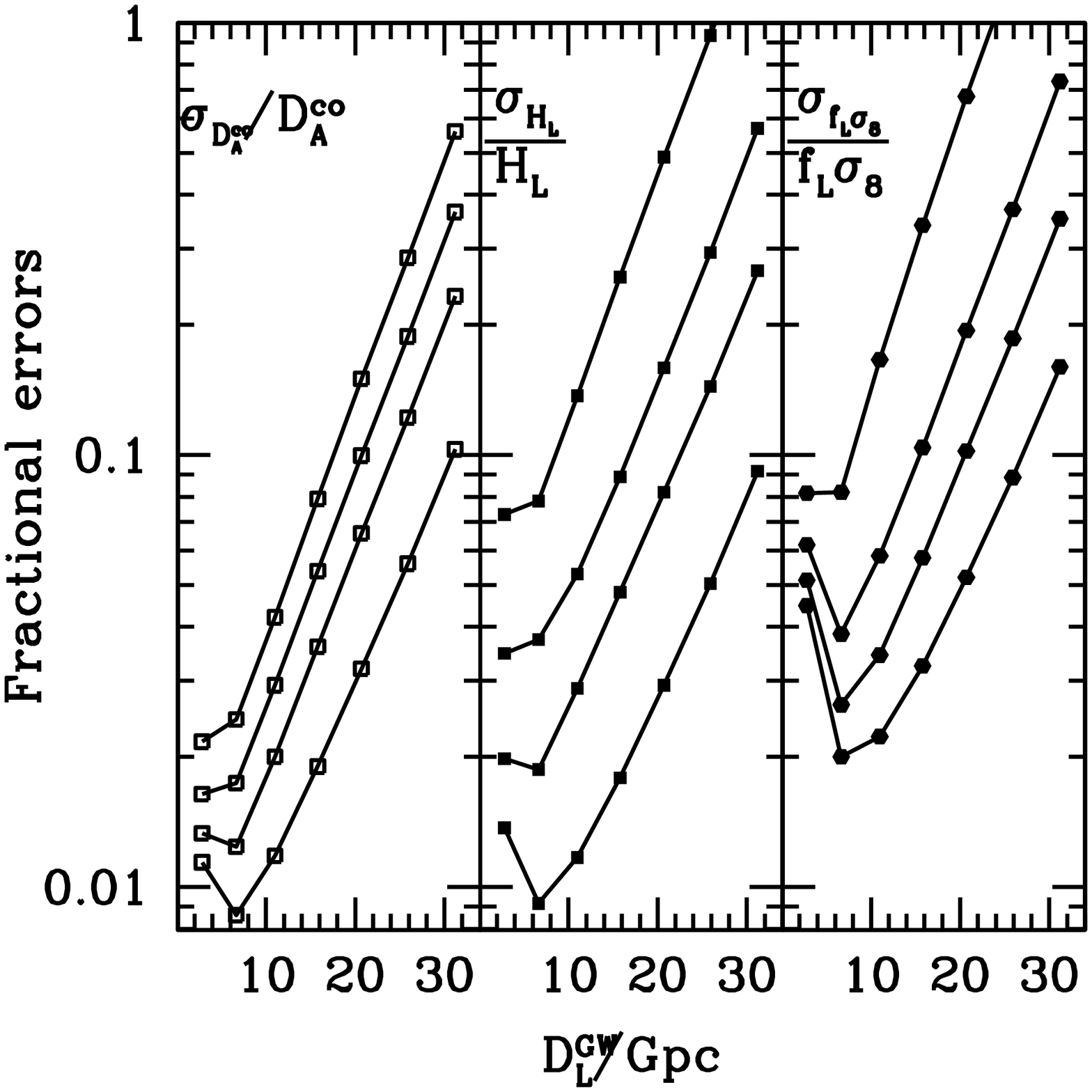}
\caption{The forecasted measurement errors on $D_A^{\rm co}$, $H_L$
  and $f_L\sigma_8$ for a number of $D_L^{\rm GW}$ bins, assuming a
  ten year observation with a BBO-like experiment. These
  measurements can constrain dark energy (Fig. \ref{fig:DL}), or determine redshift adopting the
  distance duality.  The distance
  measurement error $\sigma_D$ is a major limiting factor. BBO can reach $\sigma_{\ln  D}\sim
  0.01$ for NS-NS mergers and $\sim 0.001$ for BH-BH
  mergers. Therefore we show the cases of $\sigma_{\ln
    D}=0.02,0.01,0.005,0.001$. $\sigma_D$ degrades measurement of Fourier modes
  with $k_\parallel\ga 0.04(0.01/\sigma_{\ln D})\hmpc$. \label{fig:constraints}}
\efi

{\bf Cosmological applications}.---
Now we proceed to constraints on $D_A^{\rm co}$, $H_L$ and
$f_L\sigma_8$ using the LDS power spectrum measurement.  Assuming
Gaussian distribution in the power 
spectrum measurement errors, the Fisher matrix is 
\ba
F_{\alpha\beta}=\sum_{{\bf k}} \frac{\partial P^{\rm LDS}({\bf k})}{\partial
  \lambda_\alpha}\sigma^{-2}_{P} \frac{\partial P^{\rm LDS}({\bf k})}{\partial
  \lambda_\beta}\ .
\ea
The sum is over ${\bf k}$ bins. Instead of directly fitting $D_A^{\rm
  co}$, $1/H_L$ and $f_L\sigma_8$, we fit their ratios ($A_\perp,
A_\parallel, A_v$) with respect to the fiducial cosmology, along with
$b_g$. Namely $\lambda=(A_\perp,A_\parallel,A_v,b_g)$.   $A_\perp$ ($A_\parallel$)
scales the pair separation perpendicular (parallel) to the line
of sight. Under such scaling, 
\ba
P^{\rm LDS}(k_{\perp},k_{\parallel})\rightarrow A^{-2}_\perp
A^{-1}_\parallel P^{\rm
  LDS}\left(\frac{k_\perp}{A_\perp},\frac{k_\parallel}{A_\parallel}\right)\ .
\ea
Statistical error $\sigma_P$ in the power spectrum measurement
is
\ba
\sigma_{P}=\sqrt{\frac{2}{N_{\bf k}}}\left[P^{\rm LDS}({\bf k})+\frac{1}{\bar{n}_{\rm
  GW}}W_\parallel^{-2}(k)W_\perp^{-2}(k)\right]\ .
\ea
$N_{\bf k}$ is the number of independent Fourier modes in the ${\bf
  k}$ bin, proportional to the survey volume $V_{\rm
  survey}$. $W_{\parallel}$ ($W_{\perp}$) is the window function 
parallel(perpendicular) to the line of sight, due to statistical errors
in the $D_L$ measurement and angular positioning. 

We adopt the fiducial cosmology as the $\Lambda$CDM cosmology with
$\Omega_m=0.268$, $\Omega_\Lambda=1-\Omega_m$, $\Omega_b=0.044$,
$h=0.71$, $\sigma_8=0.83$ and $n_s=0.96$.
We are targeting at BBO or experiments of comparable capability. BBO has a
positioning accuracy better than 1 arc-minute for all NS/BH-NS/BH
mergers in the horizon \cite{2009PhRvD..80j4009C}. Since
we are only interested at large scale ($k\la 0.1\hmpc$), $W_\perp=1$
to excellent approximation. In contrast,
$W_\parallel=\exp(-k^2\chi^2\sigma_{\ln D}^2/2)$ and the
distance measurement error $\sigma_D$ has a significant effect. For typical $z\sim
1$ and $\sigma_{\ln{D}}\sim 0.01$, the induced damping is significant at
$k\ga 0.03\hmpc$. This limites the power spectrum measurement to the
linear regime. On one hand, it reduces the constraining power. On
the other hand,  it simplifies the theoretical modeling, and allows us
to neglect the FOG term in Eq. \ref{eqn:LDSD}. $\bar{n}_{\rm
  GW}$ is the average number density of GW events in the survey
volume. 
The local  NS-NS merger rate  is constrained to
$R_0=1540^{+3200}_{-1220} {\rm Gpc}^{-3}{\rm year}^{-1}$
\cite{2017PhRvL.119p1101A}.  The BH-BH merger rate is a factor of
$\sim 10$ smaller \cite{2016PhRvX...6d1015A}.  Therefore $\bar{n}_{\rm GW}$
is dominated by NS-NS mergers. For the evolution of NS-NS merger rate, we
adopt the model in
\cite{2006PhRvD..73d2001C,2009PhRvD..80j4009C}. For the bestfit $R_0$,
the total number of
GW events per year is  $0.33,1.07,1.77\times 10^6$ at $z<1,2,5$ respectively. 
\bfi{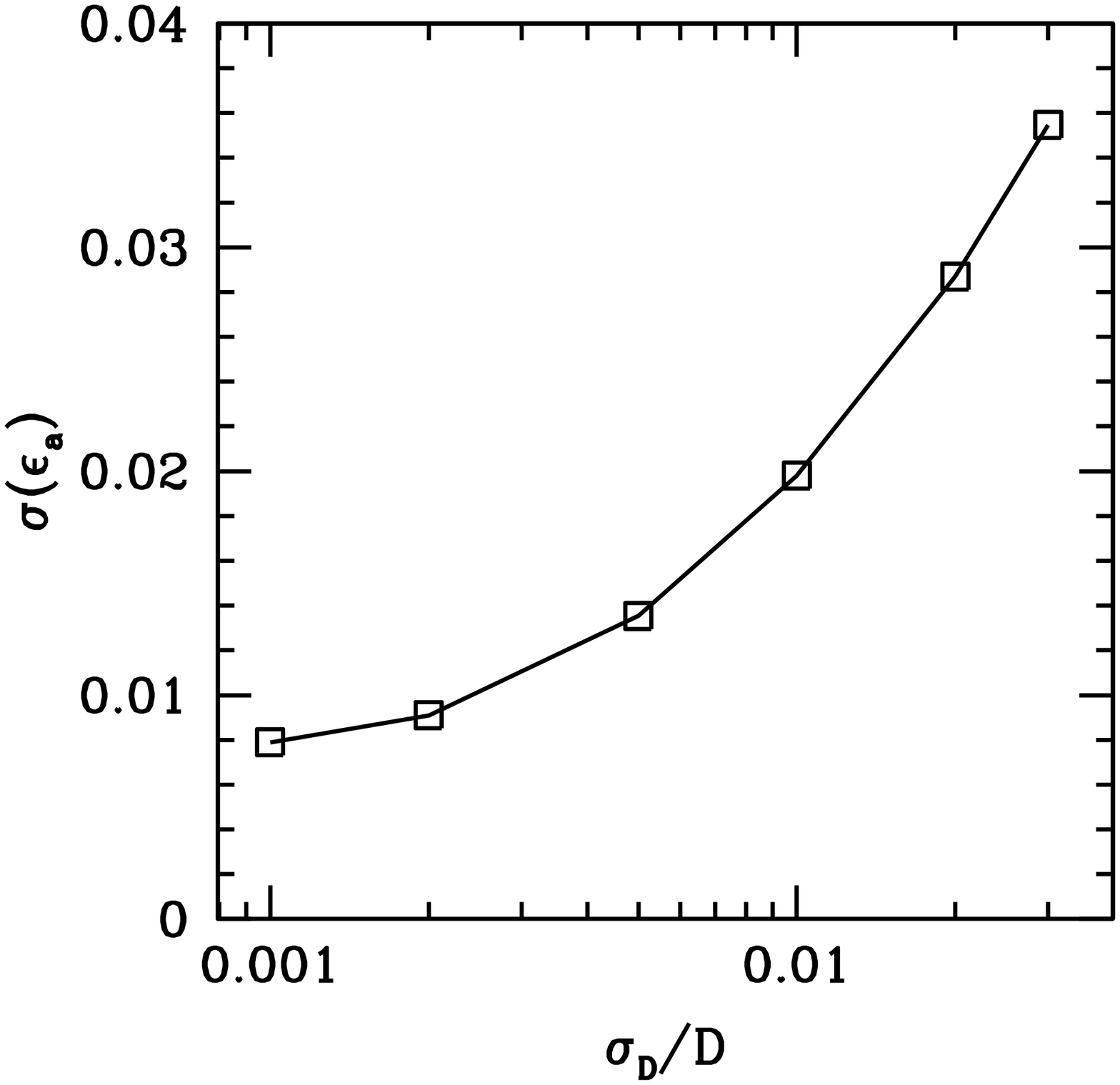}
\caption{Constraints on $D^{\rm GW}_L/D^{\rm EM}_L$, which is
  parametrized by a physically motivated parameter $\epsilon_a$.
  This will put unique and powerful constraints on modified gravity
  models. \label{fig:duality}}
\efi
We find that the
luminosity-distance space LSS is capable of constraining  $D_A^{\rm co}$, $H_L$ and $f_L\sigma_8$ in
multiple $D_L$ bins to a few percent accuracy
(Fig. \ref{fig:constraints}). These estimations adopt $\Delta t=10$ years and $b_g=1$. Since the power
spectrum measurement error is shot noise dominated, the statistical errors roughly
scale as $(R_0\Delta t)^{-1} b^{-2}_g$.  But their dependence on
$\sigma_D$ is more complicated. Fig. \ref{fig:constraints} shows the cases of
$\sigma_{\ln D}=0.005,0.01,0.02$, within the reach of BBO 
capability. $\sigma_D$ has major impact on 
cosmology, by significantly affecting the number of accessible Fourier
modes. For $\sigma_{\ln D}=0.001$ which may be achieved by BH-BH merger observations
of BBO or NS-NS mergers observations of more advanced experiments, cosmological
constraints can be significantly improved, especially for $H_L$ and
$f_L\sigma_8$. 

These constraints alone are able to constrain dark energy, demonstrated
in Fig. \ref{fig:DL}. One way to under its constraining power is that,
when the distance duality holds ($1+z=D_L/D_A^{\rm co}$), the $D_L$-$D_A^{\rm co}$ relation is equivalent to the more
familiar $z$-$D_L$ relation in the supernovae cosmology. It indeed
contains valuable information of dark energy. However, due to lower
number density and larger error in the $D_L$ measurement, these
constraints are significantly worse than what 
will be achieved by stage IV redshift surveys such as DESI \cite{DESI16} and
Euclid \cite{Euclid16}. 

Nevertheless, these measurements are unique in constraining modified
gravity (MG) models.  In these models,  GW propagation may differ from
electromagnetic wave propagation and $D_L^{\rm GW}\neq D_L^{\rm
  EM}$. This has been proposed and been applied to constrain gravity
(e.g. \cite{2007ApJ...668L.143D,2018JCAP...07..048P}).  There are two degrees of freedom to modify the GW propagation
equation \cite{2018PhRvD..97j4066B}. One allows for deviation between the
GW speed and the speed of light. However, GW170817
\cite{2017PhRvL.119p1101A} has constrained the relative difference to
be within $\mathcal{O}(10^{-15})$
\cite{2017ApJ...848L..13A}, and ruled out a large fraction
of MG models (e.g. \cite{2018PhRvL.120m1101A}). In
contrast, the other degree of freedom is essentially unconstrained. This
is to modify the friction term in the GW propagation equation.
\cite{2018PhRvD..97j4066B} parametrizes this modification as
$H(t)\rightarrow H(t)(1-\delta(t))$. To avoid confusion of  $\delta(t)$
with the commonly used LSS $\delta$ symbol, we adopt a different notation $\epsilon_{\rm
  GW}$. $\epsilon_{\rm GW}\neq 0$ leads to 
\ba
\eta\equiv \frac{D_L^{\rm GW}}{D_L^{\rm EM}}=\exp\left(-\int_0^z
  \frac{dz}{1+z}\epsilon_{\rm GW}(z)\right) \neq 1\ .
\ea
Usually we expect no deviation from GR  in the early epoch ($\epsilon_{\rm GW}(a\rightarrow 0)\rightarrow 0$). A simple
parameterization satisfying this condition is
$\epsilon_{\rm GW}(a)=\epsilon_a a$.  Under this parametrization,
$\eta=\exp(-\epsilon_a(1-a))=\exp(-\epsilon_az/(1+z))$.  

Combining the $z$-$D^{\rm co}_A$ and/or $z$-$D_L^{\rm EM}$
measurements from electromagnetic wave telescopes, and the $D_L^{\rm
  GW}$-$D_A^{\rm co}$ measurements here,  we can measure
$D^{\rm GW}_L/D^{\rm EM}_L$. Combining the $z$-$H$ and $D_L^{\rm
  GW}$-$H_L$ measurements can also constrain this ratio. BBO can
measure this ratio and constrain $\epsilon_a$ to percent level accuracy (Fig. \ref{fig:duality}).  It will then be sensitive to MG models such
as the RR model with $m^2R\Box^{-2}R$ correction in the action
\cite{2014PhRvD..89d3008M, 2018PhRvD..97j4066B}. Since this test of
gravity is on the tensor part of space-time metric, it is highly complementary to
tests on the scalar part.  The statistical error here is dominated by the
GW observations. $\sigma_{\ln D}\simeq 0.001$ will allow for better than
$1\%$ accuracy in $\epsilon_a$, and  longer observations can further help. 

{\bf Further applications}.---
We point out that future GW experiments will map LSS in a new space,
namely the luminosity-distance space (LDS), through the luminosity-distance
determined using NS/BH-NS/BH mergers. We present a proof of concept
study on its major LSS patterns (BAO and LDSD), and list a few cosmological applications (constraining
dark energy, determining cosmological redshift and probing gravity). 
It has other applications. One is to  probe the primordial
non-Gaussianity. Another is to  probe the horizon
scale gravitational potential, since it alters the luminosity distance
and generate a relativistic correction to the number density
distribution of GW events. Both require the LSS measurement near the
horizon scale. The LDS LSS is in particular
suitable since it naturally covers the whole $4\pi$ sky and can extend
to $z\gg 1$. Furthermore, the LDS LSS is free of all systematics
associated with dust extinction, star confusion, masks and survey
boundaries, due to the transparency of GWs. This will also make it
advantageous in probing horizon scale LSS. 

Including the cross correlation with the redshift space LSS
overlapping in the survey volume, its power in constraining 
cosmology can be significantly enhanced. First, it will enable
more accurate redshift determination,  in a way
independent of galaxy clustering modelling and different to existing
proposals
\cite{2016PhRvD..93h3511O,2018PhRvD..98b3502N,2018arXiv180806615M}. Since
galaxy surveys have $\mathcal{O}(10^2)$ higher number density, the
constraint on $D^{\rm GW}_L/D^{\rm EM}_L$ (and $\epsilon_a$) will be
improved by $\mathcal{O}(10)$ than what shown in
Fig. \ref{fig:duality}.  This point will be addressed in a companion paper. Combining
the LSS in the two spaces will also reduce cosmic variance in constraining primordial non-Gaussianity, 
gravitational potential and peculiar velocity, following the cosmic
variance cancellation technique \cite{McDonald09}. Furthermore, cross correlations
beween the luminosity-distance space and redshift space are also
valuable for studies of stellar evolution and galaxy formation, such
as constraining the NS/BH-galaxy relation. Notice that all these
applications only require the overlap of GW observations and galaxy
observations in cosmic volume. No spectroscopic follow-ups of GW
events are required at all. Given the advance of DESI, Euclid, SKA and
even more advance surveys \cite{Cosmicvision}, this 
requirement will be automatically satisfied.  Given these potentials,
we recommend more comprehensive studies of LSS in the
luminosity-distance space. 

Finally we address that the above proposal does not invalidate
the usefulness of spectroscopic redshift follow-ups. With spectroscopic
redshifts, the lensing field can be measured to high accuracy
\cite{2009PhRvD..80j4009C}. The velocity field, instead of the
velocity gradient causing LDSD,
can be determined as well \cite{Zhang08a}.  Therefore massive
spectroscopic follow-ups of GW events, although highly challenging,
will be highly desirable as well. 

{\bf Acknowledgement}.---
This work was supported by the National Science Foundation of China
(11621303, 11433001, 11653003, 11320101002), and  National
Basic Research Program of China (2015CB85701).


\bibliography{mybib}
\end{document}